\documentclass[12pt]{article}

\begin{document}
\begin{center}
{\bf BOSONIC FIELDS WITH TWO MASS AND SPIN STATES}

\vspace{5mm}
 S.I. Kruglov \footnote{E-mail:
krouglov@sprint.ca}\\

\vspace{5mm}
\textit{International Educational Centre, 2727 Steeles Ave. W, \# 202, \\
Toronto, Ontario, Canada M3J 3G9}
\end{center}

\begin{abstract}
We consider the bosonic fields which describe a particle which may
exist in states with spins one and zero with different masses. All
the linearly independent solutions of the equation for a free
particle are obtained in the form of the projection matrix-dyads
(density-matrices). The interaction with anomalous magnetic dipole
and quadruple electric moments of a particle is studied. The
Hamiltonian form of the first order equation has been
investigated.
\end{abstract}

\section{Introduction}

Intermediate vector bosons $W^{\pm}$, $Z^0$ and Higgs $H$ bosons
possess spin one and zero, respectively. The standard model (SM),
however, does not predict the masses of Higgs scalar particles
that is of interest at this time. The quantum field theory allows
us to construct the model of a particle which may exist in two
spin states (one and zero) with different masses [1]. Possibly
such a model can be applied for unified description of $W^{\pm}$,
$Z^0$ and $H$ bosons or other vector and scalar particles. Anyway,
it is useful to investigate such a model of fields with two spin
and mass states for different applications.

The paper is organized as follows. In section 2 we construct the
first order (11-dimensional) wave equation for particles with two
spin and mass states. Matrix form of equations is considered in
Sec. 3. All the linearly independent solutions of the equation for
a free particle will be obtained in the form of the projection
matrix-dyads in section 4. In Sec. 5 we introduce some terms in a
relativistic wave equation which describe specific interactions
with external electromagnetic fields such as anomalous magnetic
dipole and quadruple electric moments. The Hamiltonian form of an
equation is studied in Sec. 6. Section 7 contains the conclusion.
In the appendix we write out some matrices entering the equation
in the Hamiltonian form.

\section{Bosonic field equations}

A Lagrangian of general form for a charged vector field leading to
linear differential equations of not greater than second order has
the form (see [1]):
\begin{equation}
\mathcal{L}=\delta _{\mu \nu ,\sigma \rho }\left( \partial _\mu
B^*_\sigma \right) \left( \partial _\nu B_\rho \right) +\alpha
B^*_\mu B_\mu,  \label{1}
\end{equation}

where $\partial _\mu =\partial /\partial x_\mu $, $\delta _{\mu
\nu ,\sigma \rho }=a\delta _{\mu \nu }\delta _{\sigma \rho
}+b\delta _{\mu \sigma }\delta _{\nu \rho }+c\delta _{\mu \rho
}\delta _{\nu \sigma }$, in which $a$ , $b$ and $c$ are arbitrary
constants, $B^*_\mu$ is the complex conjugated field. From the
Lagrangian (1), we obtain field equations of the form
\begin{equation}
a\partial _\mu ^2B_\nu +\left( b+c\right) \partial _\nu \left(
\partial _\alpha B_\alpha \right) -\alpha B_\nu =0.  \label{2}
\end{equation}

From the freedom in the choice of the coefficients $a$, $b$, and
$c$, we can consider different cases. The requirement that $a=-1$
is necessary to have the standard kinetic term in the Lagrangian
(1). It can also be derived by the renormalization of the fields
$B_\mu \rightarrow (-a)^{1/2}B_\mu $ (the parameter $a$ should be
negative to have the positive energy of the field $ B_\mu $ ). If
$b+c=1$ and $\alpha =-m^2$, where $m$ is the rest mass, we arrive
at the Proca theory of the vector field [2]. Then it follows from
(2) that the Lorentz condition $\partial _\alpha B_\alpha =0$
occurs and there is a constraint on the field $B_\mu $. The state
with spin $0$ is excluded and there are only three degrees of
freedom which describe the spin projections $s_z=0,$ $\pm 1$.

In the general case, without any constraints, the field $B_\mu $
realizes the $\left( 0,0\right) \oplus \left( 1/2,1/2\right) $
representation of the Lorentz group and describes four degrees of
freedom which correspond to states with spins $s=0$ and $s=1$
(with three spin projections $s_z=0,$ $\pm 1$ ). To have the
Lagrangian formulation of this case when states of the field
$B_\mu $ with spins $s=0$ and $s=1$ possess the unique rest mass
$m$, we should impose on the parameters the restrictions $b+c=0$,
$a=-1$, $ \alpha =-m^2$. Then the field functions $B_\mu $ will
satisfy the Klein-Gordon-Fock equation
\begin{equation}
\left( \partial _\mu ^2-m^2\right) B_\nu =0.  \label{3}
\end{equation}

With allowance for these conditions, the Lagrangian (1) can be
rewritten as follows (within unimportant divergent-type terms):
\begin{equation}
\mathcal{L}=-\left[ \left( \partial _\mu B_\nu \right) ^{*}\left(
\partial _\mu B_\nu \right) +m^2B_\mu ^{*}B_\mu \right].  \label{4}
\end{equation}

The Lagrangian (4) can be connected also with the Stueckelberg
formulation of the vector field [3] (see also [4]). A Lagrangian
of the form (4) for real fields $B_\mu$ also was used [5] in a
gauge-invariant formulation for a massive neutral vector field.

In the general case of arbitrary parameters $a$, $b$, $c$ and
$\alpha$, we can investigate the spectre of masses of the field
$B_\mu $ with Lagrangian (1) and the equation of motion (2).
Without loss of generality, we can also set $a=-1$. It is known
that a Lagrangian is determined by the accuracy of the divergent
terms. It means that if we make the transformation of the
Lagrangian $\mathcal{L}\rightarrow \mathcal{L}+\partial _\mu
\Lambda _\mu $, where $\Lambda _\mu $ is the arbitrary function,
we will get the same equation of motion (2). Therefore only the
combination of the parameters $b+c\equiv \beta $ possesses
physical meaning. So there are two physical parameters $\alpha $
and $\beta $ in the theory. To clear up the connection of these
parameters with the masses of the field states, we work in
momentum space. Then equation (2) becomes
\begin{equation}
\left[ \left( p_\lambda ^2-\alpha \right) \delta _{\mu \nu }-\beta
p_\mu p_\nu \right] B_\nu \left( p\right) =0.  \label{5}
\end{equation}

Introducing the matrix
\begin{equation}
M=\left( p^2-\alpha \right) I_4-\beta \left( p_{\cdot }p\right),
\label{6}
\end{equation}

where $p^2=p_\lambda ^2={\bf p}^2+p_4^2={\bf p}^2-p_0^2$, the
$I_4$ is the unit 4$\times $4-matrix and $ \left( p_{\cdot
}p\right) $ is the matrix-dyad with the matrix elements $ \left(
p_{\cdot }p\right) _{\mu \nu }=p_\mu p_\nu $, it is possible to
find eigenvalues of the matrix $M$ . The matrix $M$ obeys the
``minimal" equation
\begin{equation}
\left( M+\alpha -p^2\right) \left[ M+\alpha -p^2\left( 1-\beta
\right) \right] =0  \label{7}
\end{equation}

with eigenvalues $\lambda _1=-\alpha +p^2$ , $\lambda _2=-\alpha +
p^2\left( 1-\beta \right) $. The requirement $\det M=0$ defines
the masses of the field $B_\mu $. This condition is valid if
$\lambda _1=\lambda _2=0$, which leads to
\begin{equation}
p^2=\alpha,\hspace{1.0in}p^2=\frac{\alpha }{1-\beta }. \label{8}
\end{equation}

From Eq. (8) we find the squared masses of the field $B_\mu $
corresponding to the states with spin $s=1$ and $s=0$:
\begin{equation}
m^2=-\alpha,\hspace{1.0in}m_0^2=\frac{\alpha }{\beta -1}.
\label{9}
\end{equation}

To have real masses ($m^2>0$) we arrive at the constraints:
$\alpha <0$, $ \beta <1$. At the particular case of $\beta =0$,
there is a degeneracy of mass states and the field $B_\mu $
possesses the unique squared mass $ m^2=-\alpha $. Now it is not
difficult to find first order equations which correspond to the
second order equation (5) in the coordinate representation, which
are given by
\[
\left( \beta -1\right) \partial _\mu B_\mu =B,
\]
\begin{equation}
B_{\mu \nu }=\partial _\mu B_\nu -\partial _\nu B_\mu, \label{10}
\end{equation}
\[
\partial _\nu B_{\mu \nu }-\alpha B_\mu +\partial _\mu B=0,
\]
where $B$ is a scalar and $B_{\mu \nu }$ is an antisymmetric
tensor of the second order. Inserting the first and second
equations into the third of Eq. (10), one gets
\begin{equation}
-\partial _\nu ^2B_\mu +\beta \partial _\mu \partial _\nu B_\nu
-\alpha B_\mu =0.  \label{11}
\end{equation}

Equation (11) in the momentum representation coincides with (5)
and was proposed earlier [6,7]. From (10) we find the second order
equation for the scalar field $B$:
\begin{equation}
\partial _\nu ^2B-\frac{\alpha }{\beta -1}B=0.  \label{12}
\end{equation}

It follows from (12) that the state of the field with spin $s=0$
has the squared mass $m_0^2=\alpha /\left( \beta -1\right) $ and
therefore the state of the field with spin $s=1$ corresponds to
the squared mass $ m^2=-\alpha $. In this general case when there
are no supplementary conditions, the vector field is a mixture of
quanta with the mass $m$ and spin $1$ and a quanta with spin $0$
and mass $m_0$. To have a pure spin $1$ one needs to consider the
quanta with spin $0$ as nonphysical quanta (see [6]). But in
general case one can consider a quanta with spin $0$ on the same
footing as a quanta with spin $1$.

\section{Matrix form of equations}

Let us consider the matrix formulation of the first order of the
field under consideration which is convenient for constructing the
density matrix and for some electrodynamics calculations. We will
obtain all the linearly independent solutions of the equation for
a free particle in terms of the projection matrix-dyads.

With the use of Eqs. (9) the system of field equations (10) takes
the form
\[
\partial _\mu B_\mu +\frac{m_0^2}{m^2}B=0,
\]
\begin{equation}
\partial _\nu B_\mu-\partial _\mu B_\nu+B_{\mu \nu }=0, \label{13}
\end{equation}
\[
\partial _\nu B_{\mu \nu }+\partial _\mu B+m^2 B_\mu =0.
\]

At $m_0=m$ we arrive at the massive Stueckelberg field [3]. It is
convenient to introduce the 11-dimensional function
\begin{equation}
\Psi (x)=\left\{ \psi _A(x)\right\} =\left(
\begin{array}{c}
(1/m)B(x) \\
B _\mu (x) \\
(1/m)B _{\mu \nu}(x)
\end{array}
\right) \hspace{0.5in}(A=0, \mu , [\mu \nu ]), \label{14}
\end{equation}

where $\psi_0 (x)=(1/m)B(x)$, $\psi_\mu (x)=B_\mu (x)$,
$\psi_{[\mu\nu]}(x)=(1/m)B_{\mu \nu}(x)$, $\mu$, $\nu =1$, $2$,
$3$, $4$. Let us consider the elements of the entire matrix
algebra

\begin{equation}
\left( \varepsilon ^{A,B}\right) _{CD}=\delta _{AC} \delta _{BD},
\hspace{0.5in}\varepsilon ^{A,B}\varepsilon ^{C,D}= \delta
_{BC}\varepsilon^{A,D}, \label{15}
\end{equation}

where $A$,$B$,$C$,$D=0$, $\mu$, $[\mu\nu]$. Thus, the $\varepsilon
^{A,B}$ is the $11\times11$-matrix with elements which consist of
zeroes and only one element is unity where row $A$ and column $B$
cross. Using the properties (15) equations (13) can be represented
in the form of an equation
\[
\partial _\nu \left( \varepsilon ^{\mu ,[\mu \nu ]}+\varepsilon ^{[\mu \nu
],\mu }+\varepsilon ^{\nu ,0}+\varepsilon ^{0,\nu }\right)
_{AB}\Psi _B(x)+
\]
\vspace{-8mm}
\begin{equation}
\label{16}
\end{equation}
\vspace{-8mm}
\[
+\left[ m\varepsilon ^{\mu ,\mu }+\frac{m_0^2}{m} \varepsilon
^{0,0}+\frac 12 m\varepsilon ^{[\mu \nu ],[\mu \nu ]} \right]
_{AB}\Psi _B(x)=0.
\]

Using 11-dimensional matrices
\[
\alpha _\nu =\varepsilon ^{\mu ,[\mu \nu ]}+\varepsilon ^{[\mu \nu
],\mu }+\varepsilon ^{\nu ,0}+\varepsilon ^{0,\nu },
\]
\vspace{-8mm}
\begin{equation}
\label{17}
\end{equation}
\vspace{-8mm}
\[
P_1=\varepsilon ^{\mu ,\mu }+\frac 12\varepsilon ^{[\mu \nu ],[\mu
\nu ]},\hspace{0.5in}P_0=\varepsilon ^{0,0}
\]

Eq. (16) transforms into the relativistic wave equation of the
first order:
\begin{equation}
\left( \alpha _\mu \partial _\mu +m P_1+\frac{m^2_0}{m} P_0\right)
\Psi (x)=0. \label{18}
\end{equation}

Matrices $P_1$,$P_0$ are the projection matrices (see [8]) which
obey the relations:
\begin{equation}
P_1^2=P_1,\hspace{0.5in}P_0^2=P_0,\hspace{0.5in}P_1 P_0=0,
\hspace{0.5in}P_1+P_0=I_{11}, \label{19}
\end{equation}

where $I_{11}$ is the unit matrix in $11$-dimensional space. Eq.
(18) at $ m_1=m_0$, after taking into account Eq. (19), becomes
(see [4])
\begin{equation}
\left( \alpha _\mu \partial _\mu +m\right) \Psi (x)=0. \label{20}
\end{equation}

Eq. (20) represents the Stueckelberg equation for massive fields
in the matrix form.

The matrices $\alpha _\mu $ can be written in the form
\[
\alpha _\mu =\beta _\mu ^{(1)}+\beta _\mu ^{(0)},
\]
\begin{equation}
\beta _\nu ^{(1)}=\varepsilon ^{\mu ,[\mu \nu ]}+\varepsilon
^{[\mu \nu ],\mu }, \label{21}
\end{equation}
\[
\beta _\nu ^{(0)}=\varepsilon ^{\nu ,0}+\varepsilon ^{0,\nu }.
\]

The $10-$ and $5-$dimensional matrices $\beta _\mu ^{(1)}$ and
$\beta _\mu ^{(0)}$ obey the Petiau-Duffin-Kemmer [9-11] algebra:
\begin{equation}
\beta _\mu \beta _\nu \beta _\alpha +\beta _\alpha \beta _\nu
\beta _\mu =\delta _{\mu \nu }\beta _\alpha +\delta _{\alpha \nu
}\beta _\mu. \label{22}
\end{equation}

The equations for spin-$1$ and spin-$0$ particles are represented
as [12]
\begin{equation}
\left( \beta _\mu ^{(1)}\partial _\mu +m\right) \Psi ^{(1)}(x)=0,
\hspace{0.5in}\Psi ^{(1)}(x)=\left(
\begin{array}{c}
\psi_\mu (x)\\
\psi_{[\mu \nu ]}(x)
\end{array}
\right),  \label{23}
\end{equation}
\begin{equation}
\left( \beta _\mu ^{(0)}\partial _\mu +m_0\right) \Psi
^{(0)}(x)=0, \hspace{0.5in}\Psi ^{(0)}(x)=\left(
\begin{array}{c}
\psi_0 (x)\\
\psi_\mu (x)
\end{array}
\right).  \label{24}
\end{equation}

Eq. (23) is the $10-$dimensional Petiau-Duffin-Kemmer equation
which is equivalent to the Proca equations [2] for spin-$1$
particles and the $5-$dimensional Eq. (24) is equivalent to the
Klein-Gordon-Fock equation for scalar particles. The first order
$11-$dimensional matrix eqution (18) describes fields with two
spins $0,$ $1$ with different masses, $m_0$ and $m$. Using Eqs.
(15) one can verify that the $11-$dimensional matrices $\alpha
_\mu $ (17) obey the algebra (see [13]):
\[
\alpha _\mu \alpha _\nu \alpha _\alpha +\alpha _\alpha \alpha _\nu
\alpha _\mu +\alpha _\mu \alpha _\alpha \alpha _\nu +\alpha _\nu
\alpha _\alpha \alpha _\mu +\alpha _\nu \alpha _\mu \alpha _\alpha
+\alpha _\alpha \alpha _\mu \alpha _\nu =
\]
\begin{equation}
=2\left( \delta _{\mu \nu }\alpha _\alpha +\delta _{\alpha \nu
}\alpha _\mu +\delta _{\mu \alpha }\alpha _\nu \right). \label{25}
\end{equation}

Representations of the Petiau-Duffin-Kemmer algebra (22) were
investigated in [14-16]. It should be noted that algebra (25) is
more complicated than the Petiau-Duffin-Kemmer algebra (22).

\section{Solutions of first order equations}

Let us obtain the solutions to Eq. (18) corresponding to definite
values of the energy and momentum of massive particles. Using
Fourier transformations, Eq. (18) in the momentum space converts
into
\begin{equation}
-i\widehat{p}\Psi _p=\varepsilon \left(
mP_1+\frac{m_0^2}{m}P_0\right)\Psi _p, \label{26}
\end{equation}

where $\widehat{p}=\alpha _\mu p_\mu $, $p_\mu =({\bf p},ip_0)$.
The value of $\varepsilon =1$ corresponds to positive energy and
$\varepsilon =-1$ to negative energy and ${\bf p}$ is the momentum
of a particle. To extract the states corresponding to pure spin
one it is necessary to consider ten-dimensional subspace (23). For
this we use the projection operator $P_1$ (17) with the property
(19). Acting on the left and right sides of Eq. (26) by the
operator $P_1$ we get
\begin{equation}
-i\widehat{p}^{(1)}\Psi^{(1)}_p=\varepsilon m\Psi^{(1)}_p,
\label{27}
\end{equation}
where we used the relations
\[
P_1 \widehat{p}=\widehat{p}P_1=\widehat{p}^{(1)},\hspace{0.5in}
\widehat{p}^{(1)}=p_\mu\beta_\mu^{(1)},
\]
\vspace{-8mm}
\begin{equation}
\label{28}
\end{equation}
\vspace{-8mm}
\[
 \Psi ^{(1)}_p=P_1\Psi_p=\left(
\begin{array}{c}
0\\
\psi_\mu (p)\\
\psi_{[\mu \nu ]}(p)
\end{array}
\right).
\]

Using Eq. (22) one may verify that the equality
\begin{equation}
\left(\widehat{p}^{(1)}\right)^3=p^2\widehat{p}^{(1)}  \label{29}
\end{equation}

holds. With the help of the projection operator method [8], we
obtain solutions to Eq. (27) in the form
\begin{equation}
M_{\varepsilon} =\frac{i\widehat{p}^{(1)}\left(
i\widehat{p}^{(1)}-\varepsilon m\right) }{ 2m^2}.  \label{30}
\end{equation}

The projection matrix $M_\varepsilon$ obeys the relationship
\begin{equation}
M_\varepsilon ^2=M_\varepsilon.  \label{31}
\end{equation}

Columns of the matrix $M_\varepsilon $ are eigenvectors $\Psi _p$
of Eq. (27) with eigenvalues $ \varepsilon m$. The matrix $
M_\varepsilon $ may be transformed into the diagonal form with
matrix elements containing only ones and zeroes, and the
$M_\varepsilon $ acting on the $11$-dimensional vector-column
retains components corresponding to the eigenvalue $ \varepsilon
m$.

Now we consider the case of the states with spin-zero. Let us
introduce the projection operator extracting the fifth dimensional
subspace (24):
\begin{equation}
\bar{P}_0= \varepsilon^{0,0}+\varepsilon^{\mu,\mu},\label{32}
\end{equation}
obeying the relationships
\begin{equation}
\bar{P}_0^2=
\bar{P}_0,\hspace{0.5in}\bar{P}_0\alpha_\nu=\beta_\nu^{(0)}.
\label{33}
\end{equation}
After acting on Eq. (26) by the operator $\bar{P}_0$ one obtains
\begin{equation}
\left(i\varepsilon \widehat{p}^{(0)}+
m\bar{P}+\frac{m_0^2}{m}P_0\right)\Psi^{(0)}_p=0, \label{34}
\end{equation}
where
\begin{equation}
\bar{P}=\varepsilon^{\mu,\mu},\hspace{0.5in}
\Psi^{(0)}_p=\bar{P}_0\Psi_p=\left(
\begin{array}{c}
\psi_0 (p)\\
\psi_\mu (p)\\
0
\end{array}
\right), \label{35}
\end{equation}

where $\widehat{p}^{(0)}=p_\mu \beta_\mu^{(0)}$. Introducing the
the matrix of Eq.(34)
\begin{equation}
K=i\varepsilon \widehat{p}^{(0)}+ m\bar{P}+\frac{m_0^2}{m}P_0,
\label{36}
\end{equation}
it is not difficult to verify, using Eq. (15), that the matrix $K$
obeys the ``minimal" equation
\begin{equation}
\left(K^2-\frac{m^2+m_0^2}{m}K+m_0^2\right)
\left(K^2-\frac{m^2+m_0^2}{m}K+m_0^2+p^2\right)=0. \label{37}
\end{equation}
For the case of the state with spin-zero and mass $m_0$ of a
particle the relation $p^2=-m_0^2$ is valid. Then Eq. (37)
transforms into
\begin{equation}
K\left(K-m\right)\left(K-\frac{m_0^2}{m}\right)
\left(K-\frac{m^2+m_0^2}{m}\right)=0. \label{38}
\end{equation}
With the use of the general technique [8] we find from Eq. (38)
the projection operator
\begin{equation}
N_{\varepsilon}
=-\frac{m}{m_0^2\left(m^2+m_0^2\right)}\left(K-m\right)
\left(K-\frac{m_0^2}{m}\right) \left(K-\frac{m^2+m_0^2}{m}\right)
\label{39}
\end{equation}
satisfying the equation $N_{\varepsilon}^2=N_{\varepsilon}$. The
operator (39) acting on arbitrary non-zero $11$-dimensional vector
(matrix-column) extracts the solution of Eq. (34) which
corresponds to the state with spin-$0$ and mass $m_0$.

To find the states with pure spin and spin projection we have to
construct the spin operators. It is not difficult to verify that
the generators of the Lorentz group in the $11-$dimensional space
are given by
\begin{equation}
J_{\mu \nu }=\beta _\mu ^{(1)}\beta _\nu ^{(1)}-\beta _\nu
^{(1)}\beta _\mu ^{(1)}.  \label{40}
\end{equation}

Matrices (40) act in the $10-$dimensional subspace $\left(\psi
_\mu (x) ,\psi_{[\mu \nu]}(x)\right)$ as the scalar field $\psi_0
(x)= (1/m)B(x)$ is an invariant of the Lorentz transformations. It
should be noted that matrices (40) are also generators of the
Lorentz group for the case of the Petiau-Duffin-Kemmer equation
(23). With the use of the properties (15), it is verified that the
commutation relations
\begin{equation}
\left[ J_{\rho \sigma },J_{\mu \nu }\right] =\delta _{\sigma \mu
}J_{\rho \nu }+\delta _{\rho \nu }J_{\sigma \mu }-\delta _{\rho
\mu }J_{\sigma \nu }-\delta _{\sigma \nu }J_{\rho \mu },
\label{41}
\end{equation}
\begin{equation}
\left[ \alpha _\lambda ,J_{\mu \nu }\right] =\delta _{\lambda \mu
}\alpha _\nu -\delta _{\lambda \nu }\alpha _\mu  \label{42}
\end{equation}
are valid.  Eq. (41) is a commutation relation for generators of
the Lorentz group, and relation (42) guarantees that Eq. (18) is
form-invariant under the Lorentz transformations. A Hermitianizing
matrix $\eta $ entering a relativistically invariant bilinear form
\begin{equation}
\overline{\Psi }\Psi =\Psi ^{+}\eta \Psi,  \label{43}
\end{equation}

possesses the properties [16-18,8]:
\begin{equation}
\eta \alpha _i=-\alpha _i\eta ,\hspace{0.5in}\eta \alpha _4=\alpha
_4\eta \hspace{0.5in}(i=1,2,3).  \label{44}
\end{equation}

It is verified that the $\eta$ is given by

\begin{equation}
\eta =-\varepsilon ^{0,0}+2\beta _4^{(1)2}-\varepsilon ^{\mu ,\mu
}-\frac 12\varepsilon ^{[\mu \nu ],[\mu \nu ]}. \label{45}
\end{equation}

The squared Pauli-Lubanski vector (the squared spin operator)
reads
\begin{equation}
\sigma ^2=\left( \frac 1{2m}\varepsilon _{\mu \nu \alpha \beta
}p_\nu J_{\alpha \beta }\right) ^2=\frac 1{m^2}\left( J_{\mu \nu
}^2p^2-J_{\mu \sigma }J_{\nu \sigma }p_\mu p_\nu \right) .
\label{46}
\end{equation}

The operator $\sigma ^2$ obeys the ``minimal" equation
\begin{equation}
\sigma ^2\left( \sigma ^2-2\right) =0,  \label{47}
\end{equation}

where the eigenvalues of the squared spin operator $\sigma ^2$
correspond to spin-zero, $s=0$ ($0=s(s+1)$),  and spin-one, $s=1$
($2=s(s+1)$). Thus, the fields considered describe fields
possessing two spins, $s=0$ and $s=1$. One may separate these
states with the use of the projection operators

\begin{equation}
S_{(0)}^2=1-\frac{\sigma ^2}2,\hspace{0.5in}S_{(1)}^2=\frac{\sigma
^2}2. \label{48}
\end{equation}

These operators have the properties $S_{(0)}^2S_{(1)}^2=0$,
$\left( S_{(0)}^2\right) ^2=S_{(0)}^2$, $\left( S_{(1)}^2\right)
^2=S_{(1)}^2$, $S_{(0)}^2+S_{(1)}^2=1$. According to the general
properties of the projection operators, the matrices $S_{(0)}^2$,
$S_{(1)}^2$ acting on the vector-column (the wave function)
extract states with pure spin $0$ and $1$, respectively. Let us
consider the operator of the spin projection on the direction of
the momentum $\mathbf{p}$:
\begin{equation}
\sigma _p=-\frac i{2\mid \mathbf{p}\mid }\epsilon
_{abc}p_aJ_{bc}=-\frac i{\mid \mathbf{p}\mid }\epsilon
_{abc}p_a\beta _b^{(1)}\beta _c^{(1)}, \label{49}
\end{equation}

where $\mid {\bf p}\mid =\sqrt{{\bf p}_1^2+{\bf p}_2^2+{\bf p}
_3^2}.$ The spin projection operator obeys the ``minimal" matrix
equation:
\begin{equation}
\sigma _p\left( \sigma _p-1\right) \left( \sigma _p+1\right) =0.
\label{50}
\end{equation}

The projection operators corresponding to spin projection one and
zero are

\begin{equation}
\widehat{S}_{(\pm 1)}=\frac 12\sigma _p\left( \sigma _p\pm
1\right), \hspace{0.5in}\widehat{S}_{(0)}=1-\sigma _p^2,
\label{51}
\end{equation}

where operators $\widehat{S}_{(\pm 1)}$, $\widehat{S}_{(0)}$
correspond to the spin projections $s_p=\pm 1$ and  $s_p=0$,
respectively. One can verify that the commutation relations
\[
\left[ S_{(0)}^2,\widehat{p}\right] =\left[
S_{(1)}^2,\widehat{p}\right] =\left[ \widehat{S}_{(\pm
1)},\widehat{p}\right] =\left[ \widehat{S}_{(0)},
\widehat{p}\right] =0,
\]
\vspace{-8mm}
\begin{equation}
\label{52}
\end{equation}
\vspace{-8mm}
\[
\left[ S_{(0)}^2,\widehat{S}_{(\pm 1)}\right] =\left[
S_{(1)}^2,\widehat{S} _{(\pm 1)}\right] =\left[
S_{(0)}^2,\widehat{S}_{(0)}\right] =0
\]
are valid. Now we may construct the projection matrices extracting
states with pure spin, spin projection and energy:
\[
\Delta _{\varepsilon ,\pm 1}=M_\varepsilon
S_{(1)}^2\widehat{S}_{(\pm 1)}= \frac{i\widehat{p}^{(1)}\left(
i\widehat{p}^{(1)}-\varepsilon m\right) }{2m^2}\frac 12\sigma
_p\left( \sigma _p\pm 1\right),
\]
\[
\Delta _\varepsilon ^{(1)}=M_\varepsilon
S_{(1)}^2\widehat{S}_{(0)}=\frac{i \widehat{p}^{(1)}\left(
i\widehat{p}^{(1)}-\varepsilon m\right) }{2m^2}\frac{\sigma ^2}
2\left( 1-\sigma _p^2\right),
\]
\vspace{-8mm}
\begin{equation}
\label{53}
\end{equation}
\vspace{-8mm}
\[
\Delta _\varepsilon ^{(0)}=N_\varepsilon
S_{(0)}^2\widehat{S}_{(0)}
\]
\[
=-\frac{m\left(K-m\right)}{m_0^2\left(m^2+m_0^2\right)}
\left(K-\frac{m_0^2}{m}\right) \left(K-\frac{m^2+m_0^2}{m}\right)
\left( 1-\frac{ \sigma ^2}2\right) \left( 1-\sigma _p^2\right).
\]

Here we took into consideration the relationship $\left( \sigma
^2/2\right) \sigma _p=\sigma _p$. The projection operators $\Delta
_{\varepsilon ,\pm 1}$, $\Delta _\varepsilon ^{(1)}$ correspond to
states with spin $1$ and spin projections $\pm 1 $, $0$, and the
$\Delta _\varepsilon ^{(0)}$ extracts the spin $0$. The operators
$\Delta _{\varepsilon ,\pm 1}$, $\Delta _\varepsilon ^{(1)}$,
$\Delta _\varepsilon ^{(0)}$ are also the density matrices for
pure spin spates. One can easily construct the impure states by
summation of Eqs. (53) over the spin and spin projections. In
accordance with the approach in [8], the projection operators for
pure states may be represented as matrix-dyads:
\[
\Delta _{\varepsilon ,\pm 1}=\Psi _{\varepsilon ,\pm 1}(p)\cdot
\overline{\Psi}_{\varepsilon ,\pm 1}(p),\hspace{0.1in}\Delta
_\varepsilon ^{(1)}=\Psi_\varepsilon (p)\cdot \overline{\Psi
}_\varepsilon (p),
\]
\vspace{-8mm}
\begin{equation}
\label{54}
\end{equation}
\vspace{-8mm}
\[
\hspace{0.1in}\Delta _\varepsilon ^{(0)}=\Psi
_{\varepsilon}^{(0)}(x)\cdot \overline{\Psi} _{\varepsilon
}^{(0)}(p).
\]

The wave functions $\Psi _{\varepsilon ,\pm 1}$, $\Psi
_\varepsilon $ correspond to spin-$1$ and spin projections $\pm 1$
and $0$, respectively, and the $\Psi _\varepsilon ^{(0)}$
corresponds to the spin-$0$. The density matrices (53) and (54)
may be used for calculating different electrodynamics processes
involving polarized charged particles possessing two spins, one
and zero.

\section{Electromagnetic interactions of bosonic fields}

Introducing the interaction with electromagnetic field in the
first order equation (18) by the substitution $\partial _\mu
\rightarrow D_\mu =\partial _\mu -ieA_\mu $ ($A_\mu $ is the
four-vector potential of the electromagnetic field), and adding
terms in the relativistic manner, which are linear in
$\mathcal{F}_{\mu \nu }=\partial _\mu A_\nu -\partial _\nu A_\mu
$, we arrive at the matrix equation
\begin{equation}
\biggl [\alpha _\mu D_\mu +\frac i2\left( \kappa _0P_0+\kappa
_1\overline{P} +\kappa _2\overline{\overline{P}}\right) \alpha
_{\mu \nu }\mathcal{F}_{\mu \nu }+m P_1+\frac{m^2_0}{m} P_0\biggr
]\Psi (x)=0,  \label{55}
\end{equation}

where the projection operators $P_0$, $\overline{P}$,
$\overline{\overline{P} }$ are given by

\begin{equation}
P_0=\varepsilon ^{0,0},\hspace{0.3in}\overline{P}=\varepsilon
^{\mu ,\mu }, \hspace{0.3in}\overline{\overline{P}}=\frac
12\varepsilon ^{[\mu \nu ],[\mu \nu ]},  \label{56}
\end{equation}

and
\begin{equation}
\alpha _{\mu \nu }=\alpha _\mu \alpha _\nu -\alpha _\nu \alpha
_\mu. \label{57}
\end{equation}

The projection operators $P_0$, $\overline{P}$,
$\overline{\overline{P}}$ extract scalar, vector and tensor parts
of the wave function $\Psi (x)$ respectively.

Using the definition of matrices (15) and Eqs. (55) and (56), we
arrive from Eq. (55) at the following system of equations

\[
D_\mu \psi_\mu +\frac{m_0^2}{m}\psi_0+i\kappa
_0\mathcal{F}_{\mu\nu}\psi_{\mu\nu}=0,
\]
\begin{equation}
m\psi_{\mu \nu }-D_\mu \psi_\nu +D_\nu
\psi_\mu+i\kappa_2\left(\mathcal{F}_{\nu\rho}\psi_{\mu\rho}-
\mathcal{F}_{\mu\rho}\psi_{\nu\rho}\right)=0, \label{58}
\end{equation}
\[
D_\nu \psi_{\mu \nu }+D_\mu \psi_0+m \psi_\mu
+2i\kappa_1\mathcal{F}_{\mu\nu}\psi_\nu=0,
\]
where $\psi_{\mu \nu }\equiv\psi_{[\mu \nu ]}$. To clear up the
physical meaning of constants $\kappa_0$, $\kappa_1$, $\kappa_2$
introduced we consider the second order equation for four-vector
$\psi_\mu$. For finding such an equation it is necessary to
express the antisymmetric tensor $\psi_{\mu\nu}$ via the
four-vector $\psi_\mu$. To simplify the problem we imply the
smallness of the constant $\kappa_2$. With the use of this natural
assumption, neglecting $\kappa^2_2$ and the higher degrees of the
constant $\kappa_2$ we find from the second equation of (58) the
approximate expression

\[
\psi_{\mu \nu }\approx\frac1m \left(D_\mu \psi_\nu -D_\nu
\psi_\mu\right)
\]
\vspace{-8mm}
\begin{equation}
\label{59}
\end{equation}
\vspace{-8mm}
\[
+i\frac{\kappa_2}{m^2}\left[\mathcal{F}_{\mu\rho}\left(D_\nu\psi_\rho-
D_\rho\psi_\nu\right)-\mathcal{F}_{\nu\rho}\left(D_\mu\psi_\rho-
D_\rho\psi_\mu\right)\right].
\]

It follows from Eqs. (58) using Eq. (59) the second order equation
is

\[
\left(D^2_\nu -m^2\right)\psi_{\mu}+\Delta D_\mu D_\nu \psi_\nu
-i\left(e+2\kappa_1m\right)\mathcal{F}_{\mu\nu}\psi_\nu+
i\frac{2\kappa_0 m}{m_0^2}D_\mu
\mathcal{F}_{\rho\nu}D_\rho\psi_\nu
\]
\vspace{-8mm}
\begin{equation}
\label{60}
\end{equation}
\vspace{-8mm}
\[
-i\frac{\kappa_2}{m}D_\nu\left[\mathcal{F}_{\mu\rho}\left(D_\nu\psi_\rho-
D_\rho\psi_\nu\right)-\mathcal{F}_{\nu\rho}\left(D_\mu\psi_\rho-D_\rho
\psi_\mu\right)\right]=0,
\]
where $\Delta=(m^2-m_0^2)/m_0^2$. Eq. (60) shows that the magnetic
moment of a vector state of a particle is $e/(2m)+\kappa_1$ and
gyromagnetic ratio being $1+2m\kappa_1/e$ (see [1]). So,
$\kappa_1$ is anomalous magnetic moment (AMM) of a particle. The
constant $\kappa_2$ is connected with the quadrupole electric
moment (KEM) of a field [19]. The constant $\kappa_0$ also gives
the contribution to KEM of a particle but this contribution comes
from the scalar state of a field. Indeed, we can consider the pure
spin-one field by putting $m_0\rightarrow\infty$ in (60), i.e.
when the mass of a scalar state is infinity [6]. In this case
($m_0\rightarrow\infty$) Eq. (60) transforms into
\[
\left(D^2_\nu -m^2\right)\psi_{\mu}-D_\mu D_\nu \psi_\nu
-i\left(e+2\kappa_1m\right)\mathcal{F}_{\mu\nu}\psi_\nu
\]
\vspace{-8mm}
\begin{equation}
\label{61}
\end{equation}
\vspace{-8mm}
\[
-i\frac{\kappa_2}{m}D_\nu\left[\mathcal{F}_{\mu\rho}\left(D_\nu\psi_\rho-
D_\rho\psi_\nu\right)-\mathcal{F}_{\nu\rho}\left(D_\mu\psi_\rho-
D_\rho\psi_\mu\right)\right]=0.
\]
Eq. (61) corresponds to the description of a vector particle on
the basis of Proca's equation with the additional terms involving
AMM and KEM of a particle. As a result Eq. (61) does not contain
the constant $\kappa_0$. Thus, the consideration of a field with
two spins, one and zero, allows us to introduce phenomenologically
more constants characterizing the electromagnetic interactions of
fields.

\section{Hamiltonian form of equation}

Now we consider the Hamiltonian form of Eq. (55) for particles
possessing AMM and KEM in the external electromagnetic field.

To determine the number of dynamical variables of the wave
function $\Psi(x)$, it is necessary to consider the Hamiltonian
form of Eq. (55). For this purpose we write Eq. (55) in the form
\[
i\alpha _4\partial _t\Psi (x)=\biggl [\alpha
_aD_a+mP_1+\frac{m_0^2}{m}P_0+eA_0\alpha _4+
\]
\vspace{-8mm}
\begin{equation}
\label{62}
\end{equation}
\vspace{-8mm}
\[
+\frac i2\left( \kappa _0P_0+\kappa _1\overline{P}+\kappa
_2\overline{ \overline{P}}\right) \alpha _{\mu \nu
}\mathcal{F}_{\mu \nu }\biggr ]\Psi (x).
\]

In order to separate the canonical and non-canonical parts of the
equation, we introduce the operators:
\begin{equation}
\Lambda =\alpha _4^2,\hspace{0.3in}\Pi =1-\alpha _4^2, \label{63}
\end{equation}

which obey the following relationships
\[
\Lambda \alpha _4=\alpha _4,\hspace{0.3in}\Pi \alpha
_4=0,\hspace{0.3in} \Lambda^2=\Lambda,
\]
\vspace{-8mm}
\begin{equation}
\label{64}
\end{equation}
\vspace{-8mm}
\[
\Pi^2=\Pi,\hspace{0.3in}\Lambda \Pi=0,\hspace{0.3in}\Lambda +\Pi
=1,
\]

where $1\equiv I_{11}$ is a unit $11\times 11-$matrix. Acting on
Eq. (62) with the operators $\alpha _4$ and $\Pi $, with the help
of the equalities
\[
\Lambda \overline{P}=\overline{P},\hspace{0.3in}\Lambda P_0=P_0,
\hspace{0.3in}\Pi \alpha _\mu \Pi =\Pi P_0=\Pi \overline{P}=0,
\hspace{0.3in}\Pi P_1=\Pi \overline{\overline{P}}=\Pi,
\]
one obtains
\[
i\partial _t\varphi (x)=m\alpha
_4P_1\Psi(x)+\frac{m_0^2}{m}\alpha_4 P_0\Psi(x)+eA_0\varphi
(x)+\alpha _4 \biggl [\alpha _aD_a+
\]
\vspace{-8mm}
\begin{equation}
\label{65}
\end{equation}
\vspace{-8mm}
\[
+\frac i2\left( \kappa _0P_0+\kappa _1\overline{P}+\kappa
_2\overline{ \overline{P}}\right) \alpha _{\mu \nu
}\mathcal{F}_{\mu \nu }\biggr ]\Psi (x),
\]
\begin{equation}
m\chi (x)+\Pi \alpha _aD_a\varphi (x)+\frac{i\kappa _2}2\Pi
\overline{ \overline{P}}\alpha _{\mu \nu }\mathcal{F}_{\mu \nu
}\Psi (x)=0, \label{66}
\end{equation}

where $\Lambda \Psi (x)=\varphi (x)$, $\Pi \Psi (x)=\chi (x)$.

It is seen from Eqs. (65) and (66) that the function $\varphi (x)$
is the canonical variable, and $\chi (x)$ is the non-canonical
function. To have the Hamiltonian form of Eq. (55) we should
exclude the non-canonical function $\chi (x)$ from Eq. (65). It
follows from Eq. (66) that
\begin{equation}
\chi (x)=-\left( 1+i\gamma \right) ^{-1}\left( \frac 1m\Pi \alpha
_aD_a+i\gamma \right) \varphi (x),\hspace{0.3in}\gamma
=\frac{\kappa _2}{2m} \Pi \overline{\overline{P}}\alpha _{\mu \nu
}\mathcal{F}_{\mu \nu }. \label{67}
\end{equation}

Neglecting $\kappa_2^2$ the equality $\gamma^2\approx 0$ holds,
and as a result
\begin{equation}
\left( 1+i\gamma \right) ^{-1}\approx\left( 1-i\gamma \right).
\label{68}
\end{equation}

Taking into account Eq. (68), and inserting the expression (67)
into Eq. (65), after some transformations (see the appendix), we
arrive at the Hamiltonian form of a field equation:
\[
i\partial_t\varphi(x)=eA_0\varphi(x)+\alpha_4\biggl\{\frac{m_0^2-m^2}
{m}P_0
\]
\begin{equation}
+\left(m+\alpha _aD_a+\frac{i\kappa_0}{2} P_0\alpha_{\mu \nu
}\mathcal{F}_{\mu \nu }\right)\left( 1-i\gamma \right)\left(1
-\frac{1}{m}\Pi \alpha_aD_a\right)
\end{equation}\label{69}
\[
+\frac{i\kappa_1}2 \overline{P}\alpha _{\mu \nu }\mathcal{F}_{\mu
\nu }+\frac{i\kappa_2}{2}\overline{\overline{P}}\alpha _{\mu \nu
}\mathcal{F}_{\mu \nu }\left( 1-\frac 1m\Pi \alpha _aD_a\right)
\biggr \}\varphi (x).
\]
We use here the approximate relation $\kappa_2\gamma\approx0$. It
should be noted that we imply only the smallness of the constant
$\kappa_2$, so $\kappa^2_2\approx0$, but the constants $\kappa_0$
and $\kappa_1$ are arbitrary. The wave function $\varphi (x)$ has
eight non-zero components: six components belong to states with
spin $1$ (three spin projections and two values of an energy), and
two components belong to states with spin $0$ (two values of an
energy). It should be noted that states with spins $0$ and $1$ are
not separated, i.e., it is impossible on the basis of Eq. (69) to
obtain two Hamiltonian equations for functions describing states
with spins $0$ and $1$ separately.

\section{Conclusion}

The theory of fields possessing two spin states, one and zero, AMM
and KEM can be considered as effective one. This scheme may be
applied for composite systems in particle and nuclear physics (see
[20). It should be noted, however, that the spin-zero state gives
the negative contribution to the Hamiltonian of fields under
consideration and it is necessary to introduce an indefinite
metric to quantize such a field (see [4] where the case $m=m_0$
was studied). But consideration of fields with two spins is
justified for phenomenological applications and possibly for
constructing other field schemes.

The density matrices (matrix-dyads) calculated for fields with
spins one and zero allow us to make evaluations of different
physical quantities in a covariant manner (see [8]).

\begin{center}
{\bf APPENDIX}
\end{center}

Here we obtain the matrices entering Eq. (69) via elements of
entire algebra (15). Using the definitions (15) and with the help
of the equality [12] $\delta _{[\mu \nu ][\rho \sigma ]}=\delta
_{\mu \rho }\delta _{\nu \sigma }-\delta _{\mu \sigma }\delta
_{\nu \rho }$, it is easy to find the following expressions
\[
\alpha_\mu\alpha_\nu=\varepsilon^{[\alpha\mu],[\alpha\nu]}+\delta_{\mu\nu}
\left(\varepsilon^{\alpha,\alpha}+\varepsilon^{0,0}\right)+
\varepsilon^{\mu,\nu}-\varepsilon^{\nu,\mu}
+\varepsilon^{0,[\mu\nu]}+\varepsilon^{[\nu\mu],0},
\]
\[
\frac12P_0\alpha_{\mu\nu}=\varepsilon^{0,[\mu\nu]},
\]
\[
\frac12\overline{P}\alpha_{\mu\nu}=
\varepsilon^{\mu,\nu}-\varepsilon^{\nu,\mu},
\]
\[
\overline{\overline{P}}\alpha_{\mu\nu}=\varepsilon^{[\alpha\mu],[\alpha\nu]}
-\varepsilon^{[\alpha\nu],[\alpha\mu]}+\varepsilon^{[\nu\mu],0}-
\varepsilon^{[\mu\nu],0},
\]
\[
\Pi=\frac12 \varepsilon^{[mn],[mn]}\hspace{0.5in}(m,n=1,2,3),
\]
\[
\gamma=\frac{\kappa_2}{2m}\left(\varepsilon^{[mn],[m\nu]}\mathcal{F}_{n\nu}+
\varepsilon^{[nm],0}\mathcal{F}_{mn}\right),
\]
\[
\overline{P}\alpha_{\mu\nu}\gamma=0,\hspace{0.5in}\Pi\alpha_a=
\varepsilon^{[ma],m},
\]
\[
\gamma\Pi\alpha_a=\frac{\kappa_2}{2m}\left(\varepsilon^{[km],k}
\mathcal{F}_{ma}-\varepsilon^{[am],n}\mathcal{F}_{mn}\right).
\]
We imply the summation on repeating indexes; Greek and Latin
letters take the numbers 1,2,3,4 and 1,2,3, respectively.

\end{document}